\documentclass[a4paper,twocolumn,11pt,accepted=2022-04-19]{quantumarticle}
\pdfoutput=1
\usepackage[numbers,sort&compress]{natbib}
\usepackage[utf8]{inputenc}
\usepackage{amsmath,amssymb}
\usepackage{graphicx}
\usepackage{diagbox}
\usepackage{bm}
\usepackage{braket}
\usepackage{dsfont}
\usepackage[svgnames]{xcolor}
\usepackage{subfigure}
\usepackage{quantikz}
\usepackage{braket}
\usepackage{tikz}
\usepackage{booktabs}
\usepackage{tikz-network}
\usepackage{algorithm}
\usepackage{algpseudocode}
\usepackage{algorithmicx}
\usepackage[colorlinks,linkcolor={DarkBlue},citecolor={DarkBlue},urlcolor={DarkBlue}]{hyperref}
\usepackage{cleveref}
\usepackage[activate={true,nocompatibility},final,tracking=true,kerning=true,spacing=true,factor=1100,stretch=10,shrink=10]{microtype}

\begin{document}
	\title{Jet: Fast quantum circuit simulations with parallel \newline task-based tensor-network contraction}
	\author{Trevor Vincent}
	   \affiliation{Xanadu, 777 Bay Street, Toronto, Canada}
    \author{Lee J. O'Riordan}
       \affiliation{Xanadu, 777 Bay Street, Toronto, Canada}
  	\author{Mikhail Andrenkov}
  	   \affiliation{Xanadu, 777 Bay Street, Toronto, Canada}
  	\author{Jack Brown}
  	   \affiliation{Xanadu, 777 Bay Street, Toronto, Canada}
  	\author{Nathan Killoran}
  	   \affiliation{Xanadu, 777 Bay Street, Toronto, Canada}
  	\author{Haoyu Qi}
  	   \affiliation{Xanadu, 777 Bay Street, Toronto, Canada}
  	 \author{Ish Dhand}
   \affiliation{Xanadu, 777 Bay Street, Toronto, Canada}
      \affiliation{Institute of Theoretical Physics and IQST, Ulm University, Albert-Einstein-Allee 11, 89081 Ulm, Germany}
	
\begin{abstract}
		We introduce a new open-source software library \textit{Jet}, which uses task-based parallelism to obtain speed-ups in classical tensor-network simulations of quantum circuits.
		These speed-ups result from i)~the increased parallelism introduced by mapping the tensor-network simulation to a task-based framework, ii)~a novel method of reusing shared work between tensor-network contraction tasks, and iii)~the concurrent contraction of tensor networks on all available hardware.~We demonstrate the advantages of our method by benchmarking our code on several Sycamore-53 and Gaussian boson sampling (GBS) supremacy circuits against other simulators. We also provide and compare theoretical performance estimates for tensor-network simulations of Sycamore-53 and GBS supremacy circuits for the first time. 
\end{abstract}

\maketitle

\section{Introduction}

Classical simulation of quantum systems are the workhorse of research in quantum many-body physics and quantum information processing. In particular, a large body of research exists for the classical simulations of random quantum circuits (RQCs) as they are good candidates for demonstrating quantum advantage~\cite{boixo2018characterizing,villalonga2020establishing,gray2021hyper,pednault2017breaking,huang2020classical}.~In a landmark paper, Arute et al.~\cite{arute2019quantum} demonstrated a task that was beyond the capability of current classical computers using their Sycamore-53 device on a 53-qubit RQC with a depth ($m$) of 20 cycles. However, it was argued shortly thereafter that the classical simulation benchmarks used in~\cite{arute2019quantum} were overestimates and could be greatly improved upon by using secondary storage of the Summit supercomputer~\cite{pednault2019leveraging}. Deshpande et al.~\cite{deshpande2021quantum} have proposed a quantum advantage experiment using Gaussian boson sampling (GBS) and a three-dimensional RQC. It appears these circuits could be much harder to simulate classically than Sycamore-53, with some simulations showing it would take $\approx 10^{14}$ seconds on the top supercomputer in the most idealized scenario where unlimited memory was available.
While there are many different classical methods for simulating RQCs, currently the fastest known methods involve tensor networks, with Huang et al.~\cite{huang2020classical} having the shortest estimates for simulating the Sycamore-53 device. Despite the level of complexity and innovation of the algorithms used in their classical simulations, the estimate is still orders of magnitude slower than the Sycamore-53 device.~A major issue in the simulations of~\cite{huang2020classical} was efficiency, as the simulator utilized only around 15 percent of the theoretical compute performance of the NVIDIA V100 GPU they performed the tests on.~A second issue was the lack of CPU usage in their simulations, which would result in under-utilization of the Summit supercomputer, which has 9,216 IBM POWER9 22-core CPUs amongst the 27,648 NVIDIA Tesla V100 GPUs. These issues may be accentuated as supercomputers reach exascale performance in the next few years.

Current inefficiencies in tensor-network simulation on supercomputers are likely to worsen with the oncoming era of exascale computing and the challenges it will bring. Future exascale supercomputers are expected to increase in parallelism and heterogeneity, with billions of threads running concurrently~\cite{bergman2008exascale,heldens2020landscape}. Specifically, the systems will not increase in the number of nodes, but rather in on-node concurrency with large multi-core CPUs and multiple GPUs per node~\cite{dongarra2014applied,top500website,dongarra2020report}. In addition, memory hierarchies are expected to become deeper; however, the memory size and memory bandwidth per core is expected to decrease. Due to the large number of components in an exascale machine, hardware failures are expected to
increase and fault-tolerance will be a major issue. The programming models used in tensor-network simulators will likely have problems addressing these changes. 

To address issues with current programming models, the high-performance computing community has started considering
asynchronous task-parallelism (sometimes referred to as asynchronous many-task run-time models) as a solution~\cite{thoman2018taxonomy,heldens2020landscape}.~Asynchronous task-parallelism is founded on the idea of decomposing an algorithm into units of work, known as tasks, and executing them asynchronously.~Asynchronicity allows the hiding of
communication latencies, whilst dependencies between tasks enforce program correctness
and allow for finer-grained synchronisation.~Furthermore, to solve the problem of heterogeneity, tasking would be a sensible approach as it allows for implicitly handling load balancing by executing work when needed,
as opposed to the current commonly-used methods of problem sub-division.~In addition to solving problems that will exist in upcoming exascale supercomputers, task-based parallelism will allow for more irregular and complex scientific programs to be solved, which have previously been untenable due to a lack of support for a suitable programming model. Finally, asynchronous task-parallelism can also effectively handle fault-tolerance issues without global synchronization~\cite{paul2019enabling}.
Several task-parallelism libraries are currently under active development, such as Taskflow~\cite{huang2020cpp}, Charm++~\cite{kale1993charm++}, HPX~\cite{kaiser2020hpx} and Kokkos~\cite{edwards2014kokkos}.

To date,~there has been no prior work detailing how to execute tensor-network quantum circuit simulations using a task-based parallelism framework. In this work, we introduce the open-source software Jet, which builds the task-dependency graph for tensor-network-based quantum-circuit simulations to be executed efficiently on heterogeneous nodes using the Taskflow library~\cite{huang2020cpp}.~The introduction of task-parallelism with Jet provides us with three key advantages over existing tensor-network simulators.~Firstly, decomposing the tensor-network contraction problem into tasks allows for a greater degree of parallelism because multiple tensor transposes and/or contractions can be executed concurrently. Secondly, by writing out the task graph we can take advantage of shared computation between multiple tensor-network contractions without any extra effort. Lastly, we can use Taskflow to target heterogeneous CPU+GPU nodes. 

This paper is structured as follows.~In Sec.\hspace{-.06cm}~\ref{sec:background} we overview the background needed to understand tensor-network simulators and task parallelism. In Sec.~\ref{sec:results} we describe how to contract tensor networks using task-based parallelism and shared-work reuse on heterogeneous computing nodes.~We follow this in Sec.~\ref{sec:numerical} with numerical results showcasing our methods on Sycamore-53 and three-dimensional GBS circuits, including a first-ever comparison of theoretical estimates for Sycamore-53 and GBS quantum advantage circuits.  We then conclude the paper in Sec.~\ref{sec:conclusion} with viewpoints on potential future improvements and extensions.

\section{Background}
\label{sec:background}
\subsection{Tensor Networks}

A tensor network is a countable set of tensors bonded together by contracted indices~\cite{bridgeman2017hand}.~We call the reduction of a tensor network through the summation of all contracted indices, a contraction of the tensor network. Contracting tensor networks with arbitrary structure remains at least $\#P$-hard~\cite{damm2002complexity}.~The time complexity of the contraction is heavily sensitive to the order of summations~\cite{schindler2020algorithms}, with the determination of the optimal path being an NP-hard problem~\cite{chi1997optimizing}.~Despite these difficulties, several approximate and exact methods have been developed to quickly determine quasi-optimal contractions paths and contract the tensor network, with the most common methods being tree decomposition~\cite{markov2008simulating,boixo1712simulation,lykov2020tensor} and graph partitioning~\cite{kourtis2019fast,gray2021hyper,huang2020classical}. 

\subsection{Quantum Circuits as Tensor Networks}

\begin{figure}
\begin{center}
\begin{quantikz}[ampersand replacement=\&]
\lstick{$\ket{0}$} \& \gate[2]{\quad U \quad} \& \qw\\
\lstick{$\ket{0}$} \& \& \qw
\qw
\end{quantikz}
\end{center}
\caption{A general 2-qudit quantum circuit. }
\label{fig:quantum_circuit_general}
\end{figure}
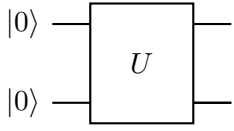

A quantum circuit is a series of logical gates acting on qudits, a $d$-level generalization of qubit. In general a quantum circuit can be represented as a unitary operator $\mathcal U$ acting on an input set of qudits each in initial state $\ket{0}$. A graphical representation of a general two-qudit circuit is shown in Fig.~\ref{fig:quantum_circuit_general}.

To demonstrate the mapping between a quantum circuit and a tensor network it is easiest to consider a concrete example of a quantum circuit, which we provide in the following example.

\subsubsection{Example Circuit}
\label{sec:example_circuit}

Consider the simple two-qudit circuit:
\begin{center}
%\begin{equation}
%\vspace{.5cm}
\begin{quantikz}[ampersand replacement=\&]
\lstick{$\ket{0}$} \& \gate{S} \&  \gate[2]{\text{B}} \& \qw \\ %\& \meter{$\ket{a_1}$}\\
\lstick{$\ket{0}$} \& \gate{S} \& \& \qw %\& \meter{$\ket{a_2}$}
\qw
\end{quantikz}
\label{eq:two_qudit_circuit}
%\end{equation}
\end{center}
Here we apply two one-qudit gates we call $S$ followed by a two-qudit gate we call $B$. The one-qudit $S$ gate can be represented by a tensor of rank 2, $S_{ba}$, whereas $B$ is a two-qudit gate represented by a tensor of rank 4, $B_{cfbe}$. Without loss of generality, we assume the initial state is the product of rank-one tensors: $\ket{0}_a\ket{0}_d$. All tensors in this example have indices with dimension $D$, the size of the qudit. We form a tensor network representation of this circuit by connecting these tensors, which represent the nodes of the network, together by the indices they share, which represent the edges of the network. For example, $S_{ba}$ is connected to $\ket{0}_a$ and $B_{cfbe}$ along the shared indices $b$ and $a$.

For classical simulations of large random quantum circuits, it is typically impossible to store the final state in memory. Therefore we can only compute amplitudes or other low-memory quantities. If we want the amplitude of the quantum state $\ket{a_1a_2}$,  we attach the rank-1 $\ket{a_1}_c$ and $\ket{a_2}_f$ tensors to the $B_{cfbe}$ tensor in the network.

The amplitude $\bra{a_1a_2}\mathcal{U}\ket{00}$, which we will shorten to  $\bra{a}\mathcal{U}\ket{0}$ hereafter, can then be computed through the contraction of a tensor network. One way of computing such a contraction is by summing over all indices in the network:
\begin{equation}
  \bra{a}\mathcal{U}\ket{0} = \sum_{abcdef}^D\bra{a_1}_c\bra{a_2}_f B_{cfbe} S_{ba} S_{ed}  \ket{0}_a \ket{0}_d,
    \label{eq:naive_summation}
\end{equation}
which has a time complexity of $\mathcal{O}(D^6)$ and is not the most optimal way to contract the network, as we'll see in the next section.

\subsection{Pairwise contraction}
Computing the amplitude by calculating the sum in Eq.~\ref{eq:naive_summation} has a time complexity of $\mathcal{O}(D^6)$  and in general has exponential cost in the number of shared indices.
It has been shown that pairwise contraction of the tensors in a tensor network is orders of magnitude faster than the naive summation in Eq.~\ref{eq:naive_summation} at computing a tensor network contraction~\cite{villalonga2019flexible,gray2021hyper,huang2020classical}. For the tensor network defined by Eq.~\ref{eq:naive_summation}, one could follow the sequence of pairwise contractions to obtain the same resulting amplitude:
\begin{equation}
    \begin{aligned}
        & 1) ~\sum_{c} \bra{a_1}_{c} B_{cfbe} = T^{1}_{fbe} \quad \mathcal{O}(D^4) \\
        & 2) ~\sum_{d}  S_{ed} \ket{0}_{d} = T^{2}_{e} \quad \mathcal{O}(D^2) \\
        & 3) ~\sum_{a} S_{ba} \ket{0}_a = T^{3}_{b} \quad \mathcal{O}(D^2) \\
        & 4) ~\sum_{f} \bra{a_2}_{f} T^{1}_{fbe} = T^{4}_{be} \quad \mathcal{O}(D^3) \\
        & 5) ~\sum_{e} T^{4}_{be} T^{2}_{e} = T^{5}_{b} \quad \mathcal{O}(D^{2}) \\
        & 6) ~\sum_{b} T^{3}_{b} T^{5}_{b} =  \bra{a}\mathcal{U}\ket{0}\quad \mathcal{O}(D).
           \label{eq:sequence}  
    \end{aligned}
\end{equation}

\begin{figure}
    \centering
    \includegraphics[width=.48\textwidth, trim = 1cm 3.5cm 1cm 1cm, clip]{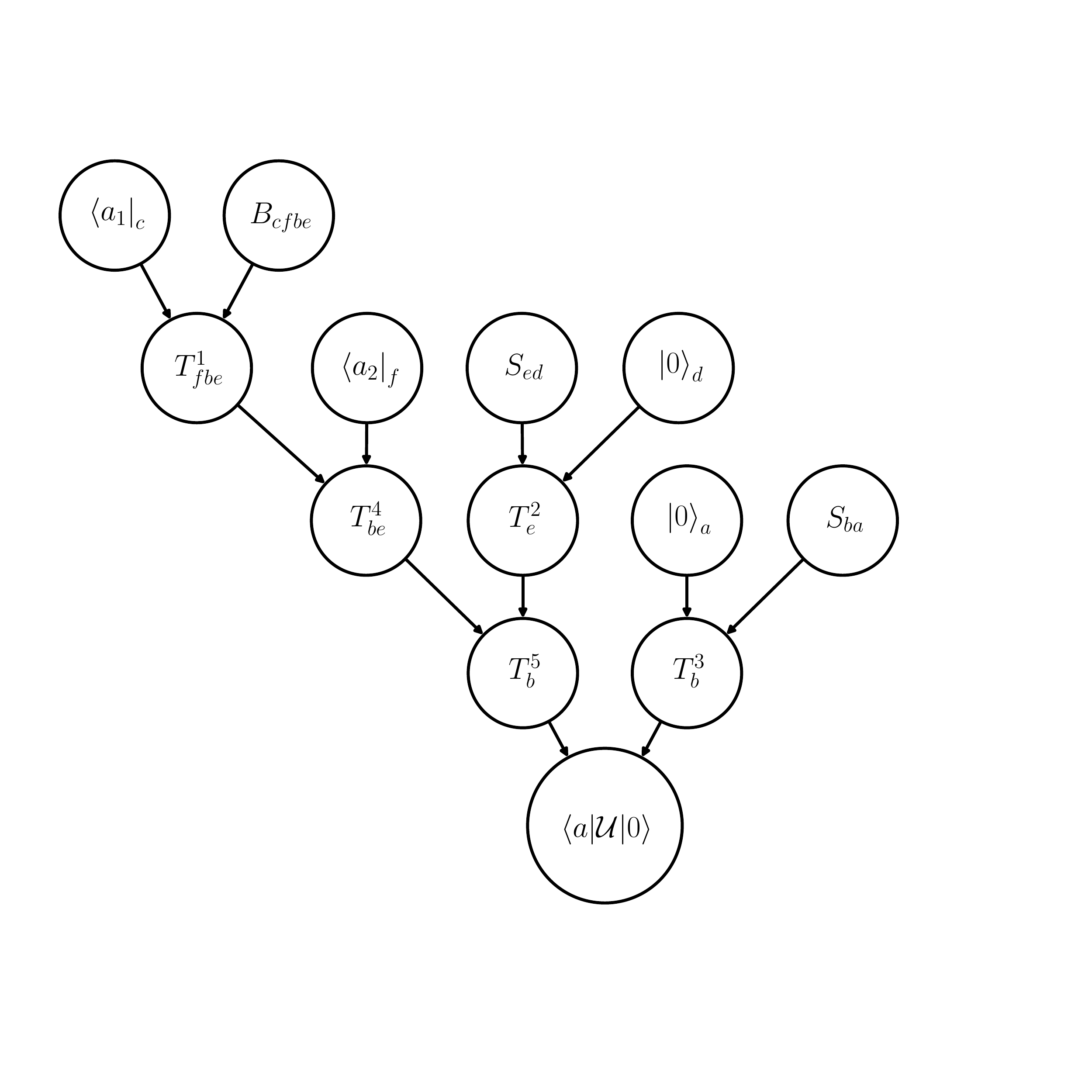} 
    \caption{The binary contraction tree for the pairwise contraction in Eq.~\ref{eq:sequence}.}
    \label{fig:contraction_tree_no_slice}
\end{figure}

The pairwise contraction in Eq.~\ref{eq:sequence} can naturally be described using a binary tree, which we show in Fig.~\ref{fig:contraction_tree_no_slice}.~Changing the contraction path will also change the binary tree associated with it.

\subsection{Slicing}

As the circuits become larger in depth and in the number of qudits, the intermediary tensors in the contraction tree grow in memory.~To avoid the need to hold the tensors in distributed memory, which would require inter-node communication to compute contractions, tensor-network simulators address the problem by using a method known as slicing or cutting \cite{villalonga2019flexible,chen2018classical,gray2021hyper}.

Slicing is the simple method of subdividing a tensor network through fixing the value of a shared index.
For example, if we consider our qudit circuit example, with the contraction defined by Eq.~\ref{eq:naive_summation} and $D=2$, then each tensor index can have a value of 0 or 1.
If we slice index $e$ in the tensor network, then Eq.~\ref{eq:naive_summation} reduces to two partial sums
    \begin{align}
        & 1)\sum_{acdfb}\bra{a_1}_c\bra{a_2}_f B_{cfb0} S_{ba} S_{0d} \ket{0}_a \ket{0}_d = s_0 \\
        & 2)\sum_{acdfb}\bra{a_1}_c\bra{a_2}_f B_{cfb1} S_{ba} S_{1d} \ket{0}_a \ket{0}_d = s_1 \\
        & 3) \sum_{i} s_i = \bra{a}\mathcal{U}\ket{0}.
        \label{eq:sliced_sum}
    \end{align}

\begin{figure}
  \centering
  \includegraphics[width=.48\textwidth, trim = 1cm 3.5cm 1cm 1cm, clip]{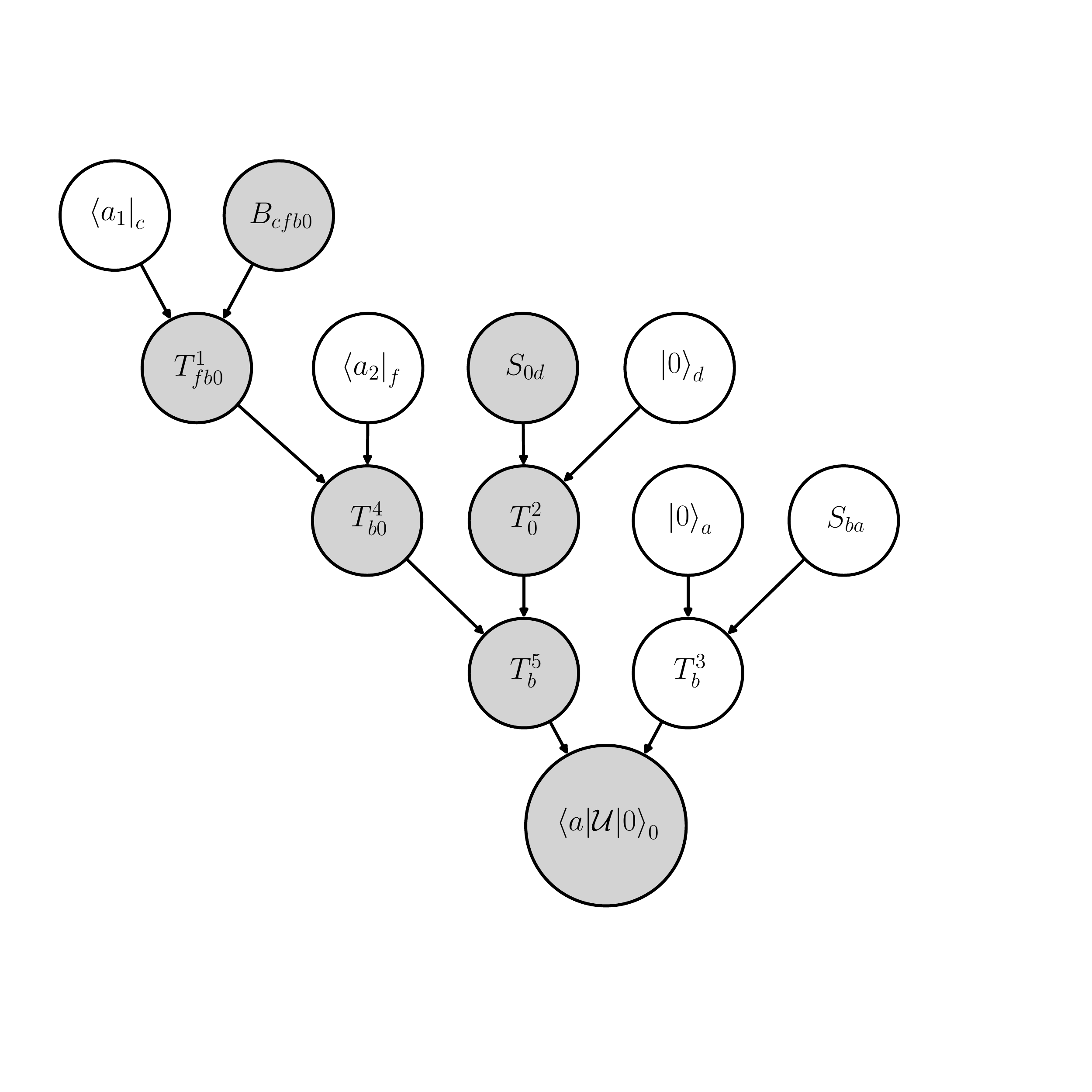}
    \caption{The binary contraction tree for a slice of Fig.~\ref{fig:contraction_tree_no_slice} through the index $e$ with value 0. We fill all nodes grey that are sliced or the result of a contraction involving a sliced tensor.}
  \label{fig:contraction_tree_slice_e}
\end{figure}

Slicing the tensor network in this manner allows for embarrassingly (data-) parallel computation, as we can distribute the two partial sums to two processing units and collect the results through a single reduction operation after the two partial sums are computed.
Equivalently, in terms of pairwise contraction, slicing the index $e$ results in two different binary contraction trees which can be distributed to two different processing units.
As an example of a sliced contraction tree, we show one of the two possible slices through the $e$ index with a value of $e = 0$ in
Fig.~\ref{fig:contraction_tree_slice_e}.
The total amount of floating point operations (FLOP) for a sliced computation of an amplitude is
\begin{equation}
\textrm{FLOP}_{\textrm{amp}} = N_{\textrm{sl}}\textrm{FLOP}_{\textrm{sl}},
\label{eq:sliced_flops}
\end{equation}
where $N_{\textrm{sl}}$ is the number of slices and $\textrm{FLOP}_{\textrm{sl}}$ is the FLOP for the contraction of a single slice.
The goal of previous tensor-network contraction algorithms was to find a set of slices and a contraction path that minimizes Eq.~\ref{eq:sliced_flops}.

Importantly, for a fixed contraction order, slicing will typically increase the total FLOP as compared to the non-sliced contraction order, known as the slicing overhead~\cite{gray2021hyper}. However, by optimizing the contraction order after slicing the network, one can often minimize or eliminate the increase in total FLOP~\cite{huang2020classical}.

\subsection{Task-based parallelism and Taskflow}

As high-performance computing (HPC) moves toward increasingly diverse architectures and larger processor counts in the exascale era, new programming models are being developed to mitigate the predicted challenges~\cite{heldens2020landscape}. The current trend of the high-performance community is a move towards asynchronous task-parallelism~\cite{mattson2016open,heldens2020landscape,bergman2008exascale,dongarra2020numerical,dongarra2014applied}, commonly referred to as asynchronous many-task parallelism. Asynchronous task-parallelism is based on the notion of a task, which represents an asynchronous operation. The size or granularity of a task can range from a single to many hardware instructions. Tasks are strung together based on dependencies to form a task graph. At run-time a scheduler continuously feeds the processing units with tasks from the task graph until the computation has completed.~Task-parallelism has demonstrated great performance for a number of complex scientific applications; see~\cite{heller2019harnessing,kidder2017spectre,9134386,phillips2020scalable,jetley2008massively,marcello2021octo}, for a few selected examples.

While there are many libraries, APIs and language extensions that implement some form of asynchronous task-parallelism, we choose the C++ library Taskflow due to its ease of use, modern design, and improved performance over other competing libraries~\cite{huang2020cpp}. Unlike many other libraries, Taskflow supports general control flow, not just a simple directed acyclic task graph, as well as a work-stealing task scheduler that can efficiently utilize heterogeneous architectures such as a multi-core CPU connected to multiple GPUs, which is common amongst supercomputer nodes. On top of this, Taskflow can handle multiple GPU vendors through the SYCL portability layer~\cite{alpay2020sycl}. While Taskflow currently only supports on-node parallelism,  we can naturally use a MPI+Taskflow model on distributed memory clusters because our problem is embarrassingly parallel due to the technique called slicing we described earlier.

\subsection{Overview of previous tensor-network simulators for quantum circuit simulation}

Many\hspace{.2cm}high-performance\hspace{.2cm}tensor-network\hspace{.2cm}simulators have been developed for various types of quantum simulation and it would be impractical to overview them all here. Instead, we limit our focus to those simulators benchmarking amplitude or sampling calculations of massive quantum circuits, such as Sycamore-53 or GBS circuits.~We also include simulators specifically developed for exascale computing. To date, there have been three major publications for tensor-network simulators benchmarking the Sycamore-53 circuit.

The first was the simulations published by Google using the tensor-network simulator qFlex for CPU simulations and TAL-SH for GPU simulations~\cite{villalonga2019flexible,arute2019quantum}. These simulations used a contraction order and set of slices predetermined by human calculation. No algorithms were used to optimize these slices or paths. The qFlex simulator had one of the fastest available CPU tensor contraction engines (specialized for tensor networks) available at the time and likewise TAL-SH was one of the fastest available for GPU tensor contraction. In Arute et al.~\cite{arute2019quantum}, they were not able to simulate the Sycamore-53 circuit to $m=20$ cycles with qFlex/TAL-SH or even provide estimates because the intermediary tensors along the contraction path could not fit on a node. Instead they computed 1 million samples to $0.5\%$ fidelity in $4.644 \times 10^3$ seconds for the $m=12$ circuit and estimated they could do 1 million samples to $0.5\%$ fidelity in $5.875 \times 10^6$ seconds for the $m=14$ circuit. This simulation was run on 4550 nodes out of 4608 of the Summit supercomputer (at the time the largest supercomputer) with 6 GPUs per node. 

The second publication came the following year with Gray and Kourtis~\cite{gray2021hyper} showcasing the strength of hypergraph partitioning methods in finding optimal contraction paths.~Together with a greedy slicing algorithm, Gray et al. dramatically reduced the runtimes for various Sycamore-53 circuits using their software library called CoTenGra~\cite{CoTenGra}. On a single NVIDIA Quadro P2000 they were able to contract an amplitude for the Sycamore-53 $m=12$ circuit in $5.74 \times 10^2$ seconds, the $m=14$ circuit in an estimated $2.92 \times 10^3$ seconds and $m=20$ in an estimated $7.17 \times 10^9$ seconds. This was the first estimate for the time complexity of the $m=20$ circuit.

The third publication came in late 2020 with Huang et al.~\cite{huang2020classical} fine-tuning the method of Gray et al. using new techniques, most notably, dynamic slicing and local optimization. Huang et al. started with a contraction order found using hypergraph partitioning and then alternated between finding a slice (using a greedy algorithm) and local optimization of the sliced contraction tree by doing exact solves of several sub-trees. Using this method in their AC-QDP simulator they ran one sample of the $m=12$ and $m=14$ circuits on a NVIDIA Tesla V100 SMX2 GPU with 16 GB of RAM. They used the results from these smaller simulations to estimate runtimes for a million samples with the appropriate fidelity on the Summit supercomputer. For $m=12,14$ they estimated it would take 18s and 88s respectively. For $m=20$ they estimated they could compute the samples in $1.67 \times 10^6$ seconds. The alterations to the Gray and Kourtis methodology introduced by Huang et al. were later incorporated into the software library of Gray and Kourtis, CoTenGra~\cite{CoTenGra}. In collaboration with NVIDIA, CoTenGra contracted a Sycamore-53 $m=20$ sample in 558 seconds on the Selene GPU cluster \cite{cuquantum}

In~\cite{deshpande2021quantum}  a three-dimensional random GBS quantum circuit containing 216 modes (6 modes per dimension of the circuit architecture) and a single cycle was proposed to showcase quantum advantage. In order to map the three-dimensional random GBS circuit to a tensor-network simulation, a truncated Fock space corresponding to a qudit size of 4 was assumed. Using the theoretical compute performance of the Fugaku supercomputer, and the CoTenGra contraction path finder~\cite{CoTenGra,gray2021hyper}, Deshpande et al.~\cite{deshpande2021quantum} found that
the 6x6x6 circuit could be computed in $\approx 10^{14}$ seconds assuming access to memory well beyond that of all the nodes of Fugaku together, let alone a single node of Fugaku. Slicing the circuit down to sizes that could fit in the 32GB RAM of a Fugaku node would come with a slicing overhead that was astronomical, making this circuit infeasible on current or future supercomputers. Ref.~\cite{deshpande2021quantum} also studied other three-dimensional GBS circuits and found that the circuit with 5 modes per dimension was also likely intractable, however an amplitude of the circuit with 4 modes per dimension could likely be computed in under an hour on Fugaku. Based on this evidence, the three-dimensional GBS circuit with 6 modes is well beyond the capabilities of current simulation methods.

Another notable tensor-network simulator is that of Lykov et al.~\cite{lykov2020tensor}, who computed 210-qubit Quantum Approximate Optimization Ansatz (QAOA) circuits with 1785 gates on 1024 nodes of the Cray XC 40 supercomputer Theta. Lykov et al. used a greedy path optimizer, which is known to perform slightly worse than the hypergraph partitioners that AQCDP and CoTenGra use~\cite{huang2020classical,gray2021hyper}. However, they coupled this greedy path optimizer with a novel step-dependent slicing method that determines indices to slice at run-time as the path is contracted. This type of method would be very suitable for dynamic task graphs and could be incorporated into the setup of Jet very naturally. Unfortunately,~\cite{lykov2020tensor} did not benchmark on Sycamore-53 so we do not know the extent to which this novel slicing method would perform against interleaving slicing with subtree-reconfiguration as was first done in ACQDP~\cite{huang2020classical} and now supported in the current version of CoTenGra~\cite{CoTenGra}. Lastly, the tensor-network simulator ExaTN~\cite{nguyen2021tensor} has been used to perform large quantum circuit simulations on Summit. Nguyen et al.~\cite{nguyen2021tensor} benchmarked ExaTN on Sycamore-53 ($m=14$), however it appears they did not perform pre-contraction simplifications or high-quality path optimization so their run-times are not competitive, despite the compute efficiency of ExaTN being quite high. 

\section{Overview of Jet}
\label{sec:results}

A major shortcoming of the simulators discussed in the previous section is that they do not take advantage of the duplicate work between slice calculations and amplitude calculations. Furthermore, they do not take advantage of the fact that some contractions in the contraction tree can be performed in parallel as they are independent. Lastly, there is no support for heterogeneous nodes and concurrent contraction on both the CPU and GPU, with the sole exception of TAL-SH and its successor ExaTN. In the following sections we introduce our simulator Jet and show how using a combination of task-based parallelism and a novel shared-work reuse we can mitigate the above difficulties.

At a high level, Jet takes in a special file describing the tensor network, which stores the contraction path and the raw tensor network data, builds a task dependency graph from this data which is then mapped to the CPU and GPU or other devices using Taskflow's task scheduler.~Our code does not perform the search for optimal slices or paths, to allow ourselves flexibility in which methods we use to compute these. Currently, to compute high-quality paths and slices for tensor networks we use CoTenGra~\cite{CoTenGra}, which uses hypergraph partitioning methods along with greedy slicing and subtree reconfiguration.~We will now overview how we build the task-dependency graph from tensor-network data step by step.

\subsection{Building the task graph for a contraction}

The basic building block of the task-dependency graph is the pairwise contraction of two tensors. This task can be decomposed into two independent (partial) tensor transposes and a single matrix multiply~\cite{lyakh2015efficient}. While tensor contractions can be computed with other approaches, the transpose-transpose-matrix-multiply method is useful in a task-based setting as it allows for more parallelism due to the fact that transposes from different contractions in the contraction tree can be executed in parallel. For contractions on the CPU we use a slightly more optimized version of the transpose used by qFlex~\cite{qFlex}, which is the fastest transpose on the CPU for quantum simulation we are aware of~\cite{villalonga2019flexible}, and we use the Intel Math Kernel Library (MKL) or OpenBLAS for the matrix multiplication. On the GPU, we use v1.3.1 of cuTENSOR~\cite{cutensor}. cuTENSOR allows us to leverage the tensor cores of the GPU when available. 

\begin{figure}[!ht]
  \centering
  \includegraphics[scale=.5, trim = .5cm 4cm 13.5cm 7cm, clip]{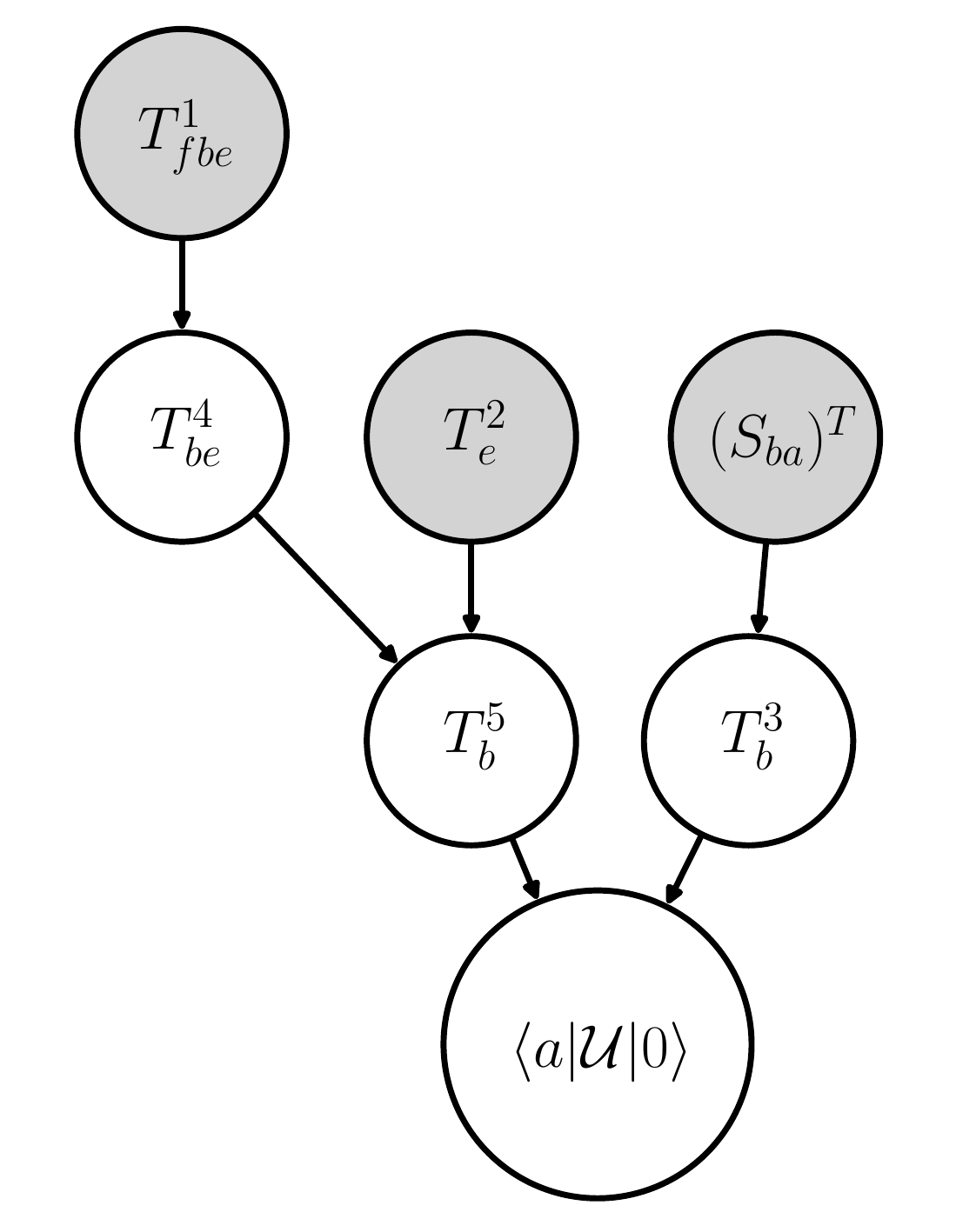}
    \caption{The task dependency graph for a single tensor-network contraction corresponding to the contraction tree in Fig.~\ref{fig:contraction_tree_no_slice}. All nodes of this task graph are named by the computational result of the task. All nodes correspond to the result of matrix-multiply calls except those with superscript $T$, which correspond to tensor transposes. In this simple contrived example, we need to transpose the tensor $S_{ba}$ because it is on the right side of the contraction between $\ket{0}_a$ and $S_{ba}$.~For this simple example, it would be possible to avoid the transpose by swapping the positions of $\ket{0}_a$ and $S_{ba}$ in the contraction tree, but in general such an operation is not possible without affecting the tensors in the rest of the tree.}
  \label{fig:single_contraction_task_graph}
\end{figure}

To create the task graph we loop through the contraction path in the tensor network file, adding the tensor transposes and matrix multiplies for each contraction in the path. Each transpose is only dependent on the tensor it is transposing being available and therefore has one dependency. The matrix multiply requires the two transposes to finish and therefore has two dependencies.
The task graph for the calculation of a single amplitude corresponding to the contraction tree in Fig.~\ref{fig:contraction_tree_no_slice} is shown in Fig.~\ref{fig:single_contraction_task_graph}.~The nodes with superscript $T$ correspond to transposes and the nodes without superscript $T$ correspond to pure matrix multiplies whose result gives the tensor named in the node.~In this simple contrived example many of the tensors do not need to be transposed before the multiplication because their indices are already in the correct order, so there are far fewer transposes than would occur in a more realistic example. Regardless, this example clearly shows that nodes highlighted in grey correspond to tasks which can be computed in parallel.  In previous simulators, which did not use task-based parallelism, taking advantage of this parallelism would not be possible.

\subsection{Building the task graph for multi-contractions and reusing shared work}

\begin{figure}[!ht]
  \centering
  \includegraphics[width=.47\textwidth, trim = 1cm 4cm 1cm 7cm, clip]{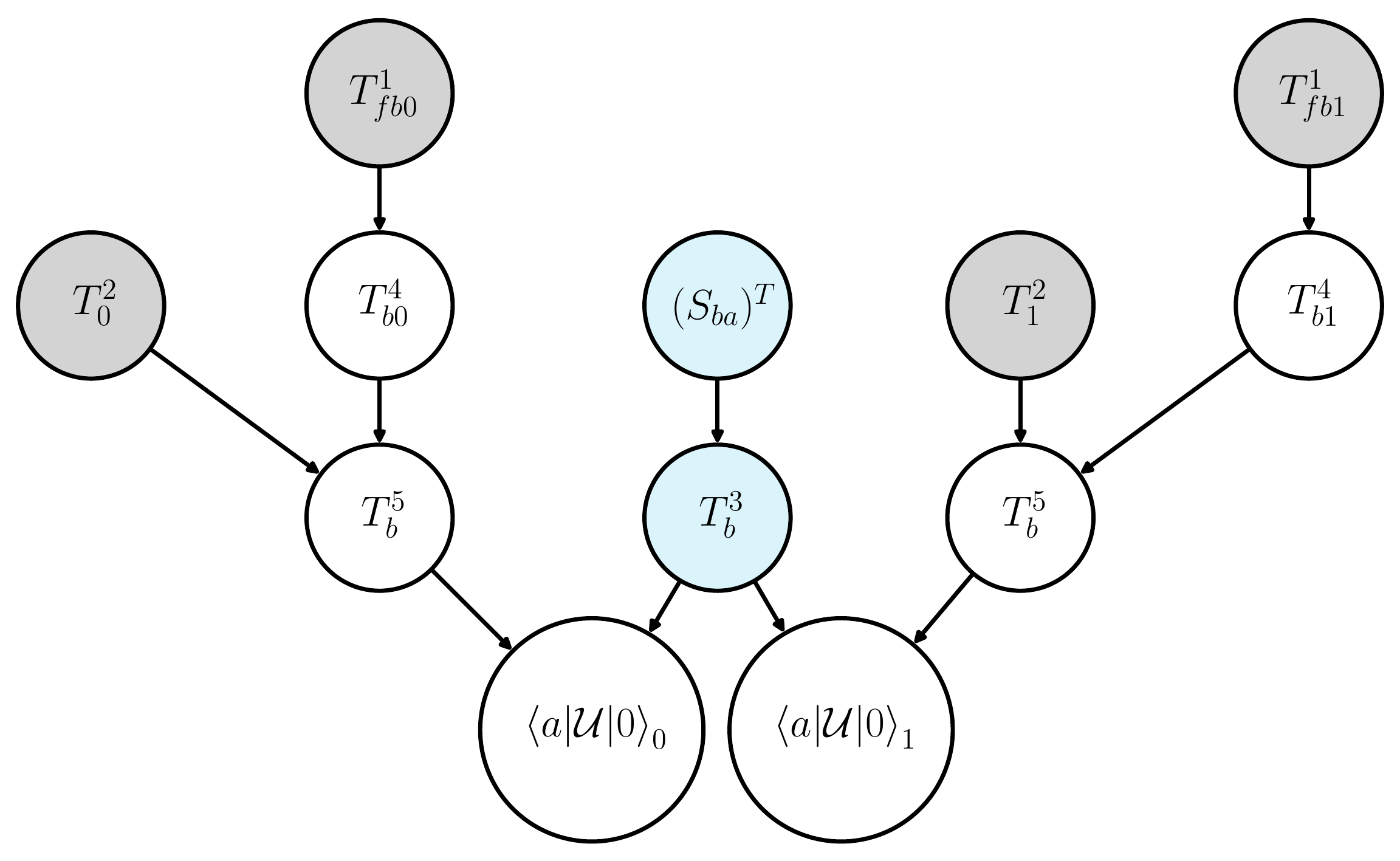}
    \caption{The task dependency graph for a multi-contraction of two slices corresponding to the fixed values of the index $e$: $e = 0$ and $e = 1$. All the filled nodes that are independent can be computed concurrently. Furthermore, the nodes colored light blue represent shared work between the two contractions. In previous simulators the light blue work would be computed twice, once for each slice. However, by using tasking we only need to compute it once.}
  \label{fig:gbs_sliced_e}
\end{figure}

In quantum circuit simulation almost all computations typically require multi-contractions, that is, multiple tensor networks will need to be contracted in order to get some final quantity of interest. Often these contractions will share work. The simplest example of this is
for computing slices.~Slicing breaks a single contraction problem into a multi-contraction problem. For large circuits, there will always be shared work between these sliced contractions. To build the task graph for this problem we loop through the contraction path for each of the slices, ignoring redundant calculations by not adding tasks with the same name.~We name the tasks by concatenating their contraction path order number, with their indices and the slice value associated to that index. This naming scheme avoids collisions when adding multiple slices to the same task graph.~We showcase the task graph in Fig.~\ref{fig:gbs_sliced_e}.~Two things are immediately of note.~Firstly, the task graph now contains more filled nodes, meaning there are more independent computations that can be contracted concurrently. Secondly, the nodes filled with a light blue color are shared between the two slice contractions. This means that the blue work only needs to be computed once and then can be reused for the other slice. If we were to compute more slices, we could keep re-using this work.

Previous simulators did not take advantage of this duplicate work, but with task-based parallelism it is not only possible, it is straightforward. We can actually take this a step further and start maximizing the amount of shared work between multi-contractions. Thus, the goal of path and slice finding within a task-based parallelism framework is then to minimize the amount of FLOP per contraction while maximizing the amount of duplicate tasks (shared work) between contractions.

The equation that should be minimized is not Eq.~\ref{eq:sliced_flops} but the following:
\begin{equation}
    \textrm{FLOP}_\textrm{amp} = f_{\textrm{sl}}\textrm{FLOP}_{\textrm{sl}} + N_{\textrm{sl}}(1-f_{\textrm{sl}})\textrm{FLOP}_{\textrm{sl}},
    \label{eq:task_based_amplitude_flops}
\end{equation}
where $f_{\textrm{sl}}$ is the fraction of $\textrm{FLOP}_{\textrm{sl}}$ that contains duplicate tasks between slices.~This generalizes to other forms of multi-contraction, such as computing batches of amplitudes and various methods of sampling quantum circuits which always require multi-contractions.~The shared work can also be stored onto disk for future computation.

\section{Numerical Results}
\label{sec:numerical}

In this section we benchmark Jet's task-based approach, and offer a comparison with existing simulators. All benchmarks can be reproduced with scripts provided in the open-source Jet repository \cite{jet}.~While there are many interesting circuits and computations that require tensor-network contractions, we restrict ourselves to contractions of random amplitudes or slices of Sycamore-53 and GBS circuits, which have both been used or proposed for quantum advantage tasks. For the Sycamore circuits, we use the circuit files provided in the attached data to Ref.~\cite{arute2019quantum}; these circuits are also provided in the CoTenGra repository~\cite{CoTenGra}. To generate the high-dimensional random GBS circuits we use the algorithm provided in Alg.~\ref{alg:gbs}.

\begin{figure}[!ht]
  \begin{algorithm}[H]
    \caption{
      Random n-dimensional GBS~circuit.\\
      The set of parameters dim, width, modes, cycles and $r$ are user-defined. We use q(i) to represent the $i$-th qumode. S and BS are the squeezer and beam splitter gates respectively. The vertical bar $|$ represents the operation of a gate on a qudit or set of qudits.
    }
    \begin{algorithmic}[1]
     \Procedure{RQC}{}
        \For{$i \gets 1$ to dim}
        \State llens[$i$] = (width)$^{i-1}$ 
         \EndFor
         \For{$k \gets 1$ to modes}
          \State  $S(r)$ $|$ $q(k)$
        \EndFor
            \For{$c \gets 1$ to cycles} 
            \For{$l \gets 1$ to dim}
            \For {$i \gets 1$ to modes$-$llens[$l$]}
                       \State $\theta$ = random$(0,2\pi)$
              \State $\phi$ = random$(0,2\pi)$     
              \State BS$(\theta,\phi)$ $|$ $q(i),q(i+\text{llens}[l])$
              \EndFor
              \EndFor
              \EndFor
      \EndProcedure
    \end{algorithmic}
      \label{alg:gbs}
  \end{algorithm}
\end{figure}

In Tab.~\ref{tab:fugaku_benchmarks} we showcase run-time estimates for a random amplitude contraction of a variety of circuits on the fastest supercomputer, Fugaku~\cite{dongarra2020report}. To find the run-time estimates, we used the contraction path finder CoTenGra~\cite{CoTenGra,gray2021hyper} and ran 200 searches each for an hour, using the KaHyPar hypergraph partitioner~\cite{schlag2016k} as the search method with default CoTenGra settings for slicing and sub-tree reconfiguration for a max tensor size of $2^{27}$. More details of these settings can be found in the reference~\cite{CoTenGra,gray2021hyper}. We re-ran this set of searches several times to test for any fluctuations in our fastest estimates, and we found that the fluctuations were within a factor of 2 or less. We show the fastest runs in the table. The code name Sycamore-53-m$x$ stands for Sycamore-53 to $x$ cycles, GBS-$zzz$-m1 stands for a three-dimensional $z\times z\times z$ GBS circuit to 1 cycle. For all the GBS circuits we assume a qudit size of 4, which is likely unrealistic for accurate simulation~\cite{deshpande2021quantum} and therefore will only serve as a lower bound on the actual computation time for an accurate amplitude calculation. One obvious outcome of the results in this table is the fact that GBS-444-m1 is only around an order of magnitude more difficult than Sycamore-53-m20 for random amplitude computation. 

\begin{table}
\centering
\begin{tabular}{cccc}\toprule
Circuit & Slices & Max Size & Time (s) \\ \midrule
%Sycamore-53-m10 & $0$ &  $6.71 \times 10^7 $ & $3.04 \times 10^{-8}$\\
Syc-53-m20 & $2^{25}$ & $2.7 \times 10^{8}$ & $3.5 \times 10^1$ \\
Syc-53-m20 & $0$ & $9.0 \times 10^{15}$ & $1.6 \times 10^1$ \\ 
GBS-444-m1 & $4^{14}$ & $6.7 \times 10^7$ & $3.6 \times 10^{2}$ \\
GBS-444-m1 & $0$ & $4.4 \times 10^{12}$ & $1.7 \times 10^{-1}$ \\
GBS-666-m1 & $0$ & $7.9 \times 10^{28}$ & $2.1 \times 10^{14}$ \\ \bottomrule
\end{tabular}
\caption{Estimated runtimes on the Fugaku supercomputer to contract one amplitude for various Sycamore-53 circuits and the three-dimensional GBS circuits GBS-444-m1 and GBS-666-m1, which have qudits arranged in 4x4x4 and 6x6x6 lattices and we apply gates to depth 1 (m=1). For the GBS circuits we use a qudit size of 4. The time estimate assumes we can compute contractions at the LINPACK benchmark speed for Fugaku, which is 442 PetaFLOP/s \cite{top500website} and that the contraction path FLOP is a good estimate of the computational work. We showcase estimates that slice no indices (see the rows with slices = 0); these estimates typically would be lower bounds on the possible run-time, but they are unrealistic because they would need more memory than is available on any supercomputer. We also showcase estimates that searched for slices which constrain the rank of intermediary tensors to be below or equal to 27, as contraction trees with this bound on tensor rank usually fit in the memory of a GPU or CPU on a supercomputer node. All estimates assume single-precision complex numbers.}
\label{tab:fugaku_benchmarks}
\end{table}

\begin{figure}[!ht]
    \centering
    \includegraphics[width=.48\textwidth]{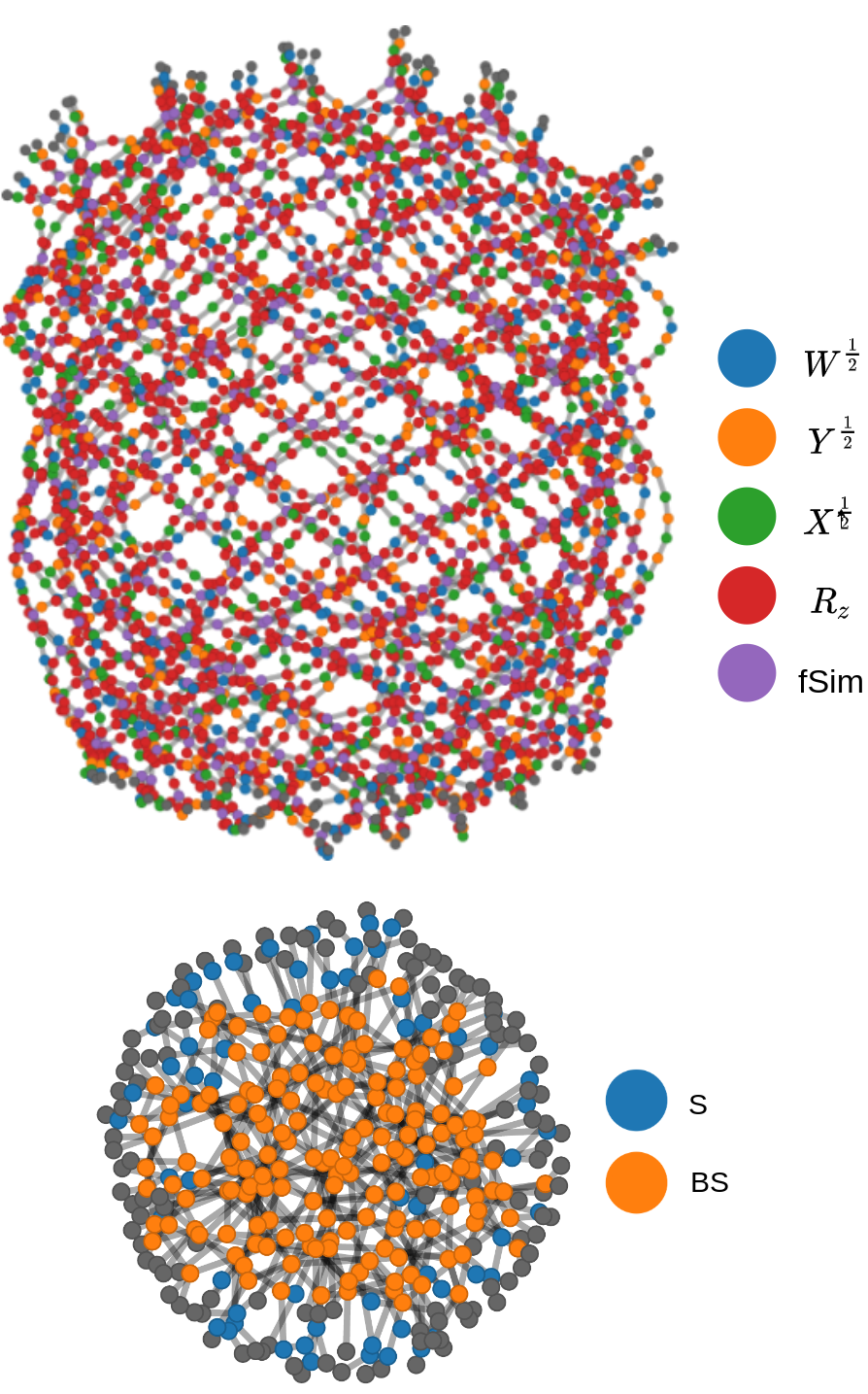}
    \caption{ {\bf top}: A plot of the tensor-network for the Sycamore-53 $m=20$ circuit. The only two qubit gate is fSim. {\bf bottom}: A $4\times4\times4$ GBS circuit with beam splitters (BS) which act as two-qudit gates and squeezers (S), which act as one-qudit gates.}
    \label{fig:tn_plots}
\end{figure}

In order to benchmark Jet's task-based parallelism, we run it on the SciNet \cite{loken2010scinet} clusters Niagara \cite{niagara,ponce2019deploying} and Rouge \cite{rouge} for our CPU benchmarks, and an AWS EC2 P4d virtual instance for our GPU benchmarks. Niagara is a large cluster of 2,024 Lenovo SD530 servers with 40 Intel Skylake cores at 2.4 GHz each (with 20 cores per socket). The Rouge cluster is a single-socket AMD Epyc 48 core processor at 2.3 GHz each. The AWS instance has two Cascade Lake 24-core processors at 3 GHz, with 8 NVIDIA A100 GPUs each having $40$ GB of device memory. For all GPU benchmarks we restrict ourselves to a single device. 

\begin{figure}
    \centering
    \includegraphics[width=.48\textwidth]{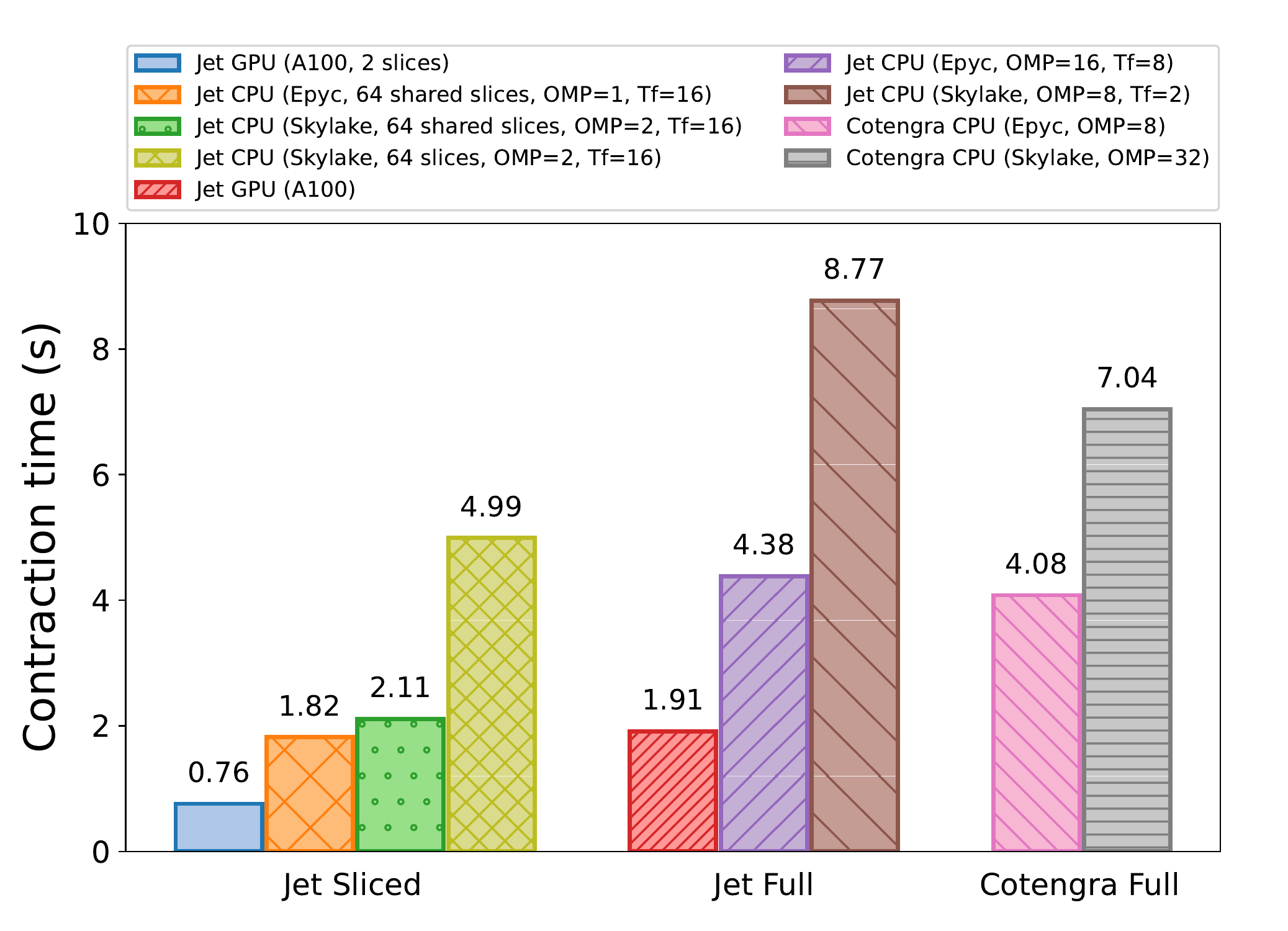}
    \caption{Jet $m=10$ contraction times for a single amplitude for the full network, and for the sliced network.~For comparison, we also show CoTenGra contraction times for the full network contraction on the available hardware.~The OMP values correspond to the optimal OpenMP threads for BLAS operations and parallelised code-paths, and Tf to the Taskflow concurrent threads for the task-graph. We can see significant benefit introduced by slicing the network and using shared work, which in the case of the CPU gives approx. $2.4\times$ and $4\times$ speedup on each respective platform, and approx. $2.5\times$ for the GPU.}
    \label{fig:m10_runtimes}
\end{figure}

In Fig.~\ref{fig:m10_runtimes} we showcase benchmarks for the computation of one amplitude of Sycamore-53-m10 across a variety of different systems, with the parameters defining the optimal minimum contraction time for the given problem and hardware, averaged over 10 runs. We can see that Jet is comparable to CoTenGra for CPU-based tensor contractions, where we show the work on a full network contraction using both Intel and AMD CPUs. In addition, we demonstrate the effect of slicing for performance, as well as using GPUs. From here we can see that the performance of the contractions are drastically improved by slicing the network, due to the inherent parallelism offered by the method. For the AMD Epyc system, slicing 6 edges (64 slices), and using 1 OpenMP thread and 16 concurrent Taskflow threads reduced the runtime by over a factor of 2. For the Skylake node, the improvement was almost a factor of 4, using 2 OpenMP threads and 16 concurrent tasks. Additionally, for the A100 GPU example, we see approximately a 2.5-fold difference. This improvement in performance can be attributed to several factors:

\begin{enumerate}
    \item By slicing the network, we may fill the task graph with work for individual slices, to be operated on in parallel, and reduce the results once all contractions are completed. This embarrassingly parallel approach allows for the exploitation of multiple-cores/devices.
    \item The task-based framework allows for easy control of the number of concurrent threads, enabling the above tasks to be consumed as the cores become available.
    \item Our contraction operations are largely a combination of BLAS-3 calls (compute bound) and data permutations (memory-bandwidth bound). Slicing reduces overall tensor sizes, and hence allows more data through the memory bus during permutations, and hence reduced BLAS-3 overhead, enabling better performance all around. The runtime difference between Skylake and Epyc nodes is likely attributed to the memory bus differences between these hardwares.
    \item The tasks resulting from slicing that have shared work with other slices, and enable a reduction in the overall operations required to contract the network.
\end{enumerate}

 For the $m=10$ dataset, we greedily selected our slices along a fixed contraction path to maximize shared work as much as possible. Our chosen contraction path and set of slices achieved a saving of 45.2 GFLOP (out of a total of 97.2 GFLOP) through use of shared work. We also save computation time because the shared transpose tasks are also no longer needed. In total, our use of shared work provides a $\approx47\%$ saving in FLOP which is reflected in the more than two-fold speedup between the sliced Skylake run that used shared work (the green bar with 2.11s in Fig.~\ref{fig:m10_runtimes}) and the sliced Skylake run that did not use the shared work (the yellow bar with 4.99s in Fig.~\ref{fig:m10_runtimes}). Whether we can get the same savings in FLOP for larger circuits or tensor-networks with different geometry would require having a more accurate method of finding slices and contraction paths that maximize shared work, which is outside the scope of this work.

\begin{figure}
    \centering
    \includegraphics[width=.48\textwidth]{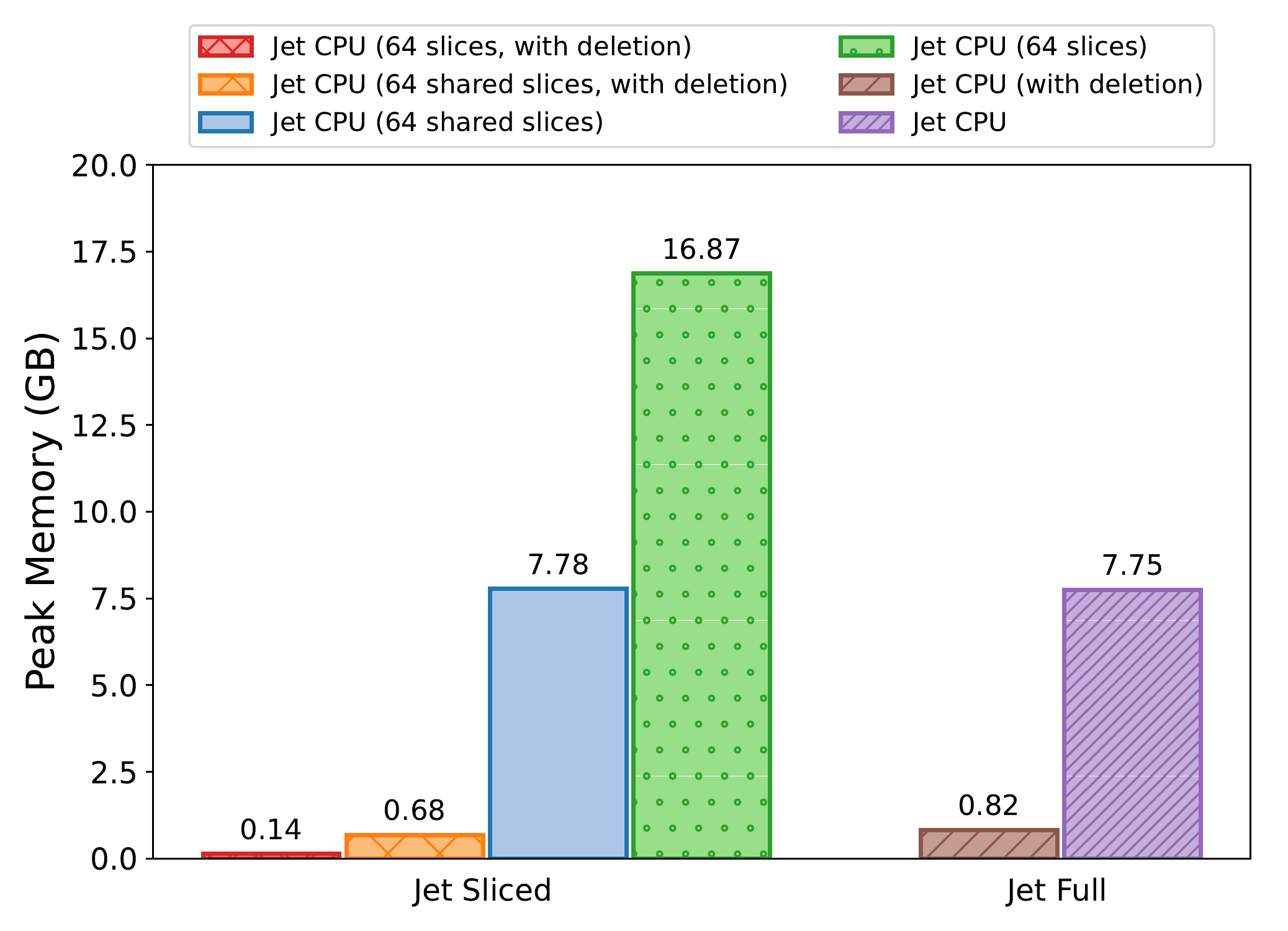}
    \caption{Jet $m=10$ peak memory consumption in gigabytes (GB) for a single amplitude of the full network and the sliced network. Each simulation is run on a single Skylake node using one thread. The bars labelled "with deletion" correspond to runs where intermediary tensors are deleted after they are no longer required.}
    \label{fig:m10_memory}
\end{figure}

In Fig.~\ref{fig:m10_memory} we showcase the peak memory consumption of the m=10 runs on a Skylake node. For each of the three different runs, we plot two bars, one bar for the contraction without deletion of intermediary tensors and a second bar for the contraction with deletion. From Fig.~\ref{fig:m10_memory}, we see that deleting intermediary tensors results in drastic savings in memory. In the case of the sliced run that does not use shared work, we see a decrease in the memory consumption by a factor of $\approx120$. Secondly, we notice that the sliced run that uses shared work (the .68 GB orange bar in Fig.~\ref{fig:m10_memory}), requires more memory consumption than when the shared work is not used (the .14 GB red bar in Fig.~\ref{fig:m10_memory}). This is due to the extra shared intermediary tensors that are stored in memory, which is a minor drawback of using shared-work when slicing. For tensor-networks bigger than the one associated with the $m=10$ circuit, the amount of slices will grow so that only a subset can be held in memory at a given time. For such cases we can determine a priori the size of this subset. Thus, we do not run into difficulties with this task-based approach as the tensor network grows larger.

\begin{figure}
    \centering
    \includegraphics[width=.48\textwidth]{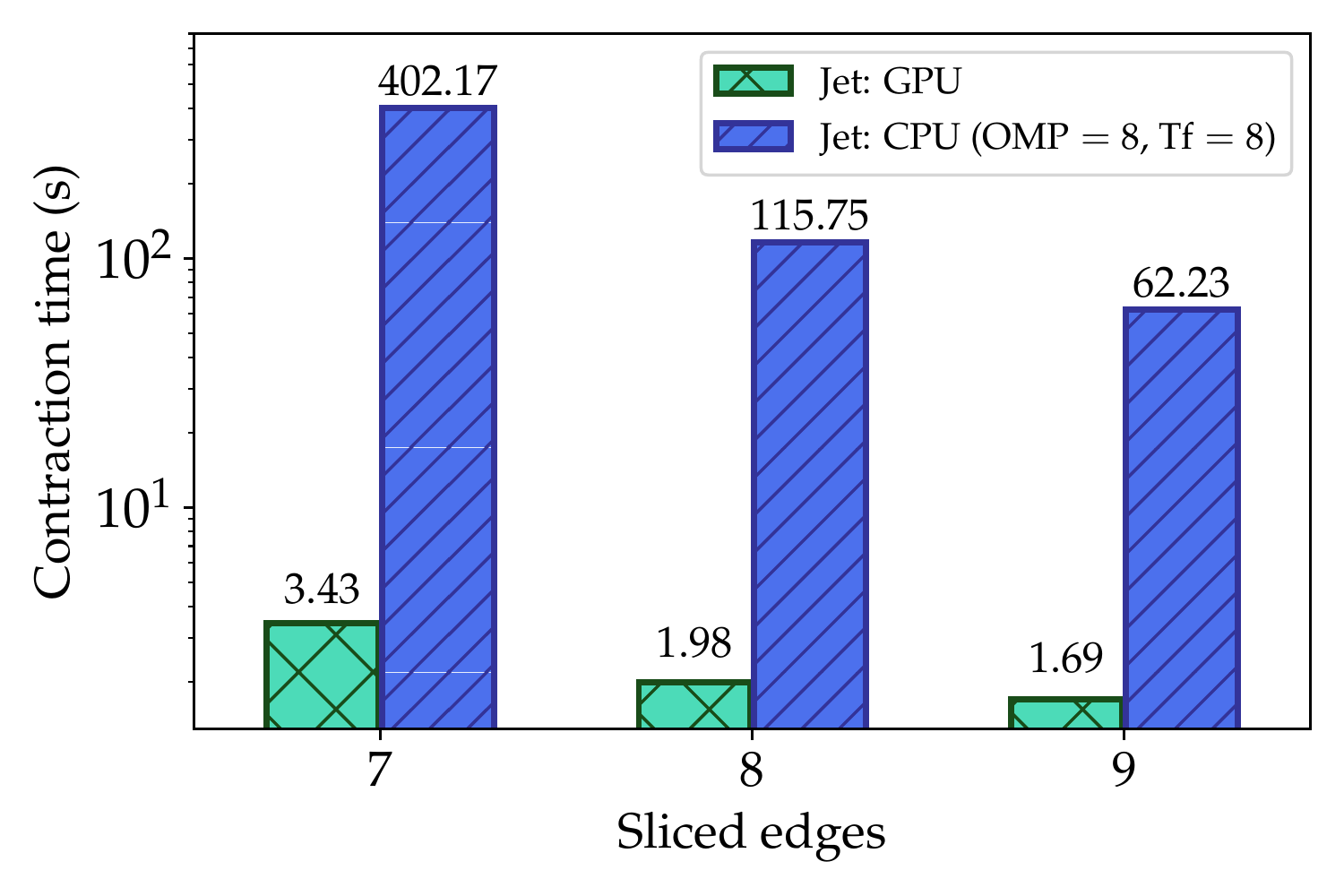}
    \caption{Runtime performance of Jet using CPU and GPU backends to contract a single slice of $m=12$. The number of sliced edges determines the sliced width, with lower numbers requiring a larger amount of data to be contracted. The CPU code ran using 8 OpenMP threads and 8 concurrent Taskflow threads, giving the optimal performance on a Niagara Skylake node. The GPU code ran on an NVIDIA A100 node with 40GB of on-board memory. Both codes used 32-bit floating point values for the complex components (\texttt{complex<float>} and \texttt{cuComplex}).}
    \label{fig:m12_jet_cpu_GPU}
\end{figure}

We also examine the use of Jet for single slices of $m=12$, comparing our CPU and GPU backends. Fig.~\ref{fig:m12_jet_cpu_GPU} showcases a significant performance difference between running Jet on an NVIDIA A100 GPU and a node of the Niagara supercomputer (Skylake CPU). For the optimal runtime parameters, the GPU is approximately 2 orders of magnitude faster than the CPU, and showcases the gains one can make using GPUs for these contraction problems. 

\begin{figure}[!ht]
    \centering
    \includegraphics[width=.48\textwidth]{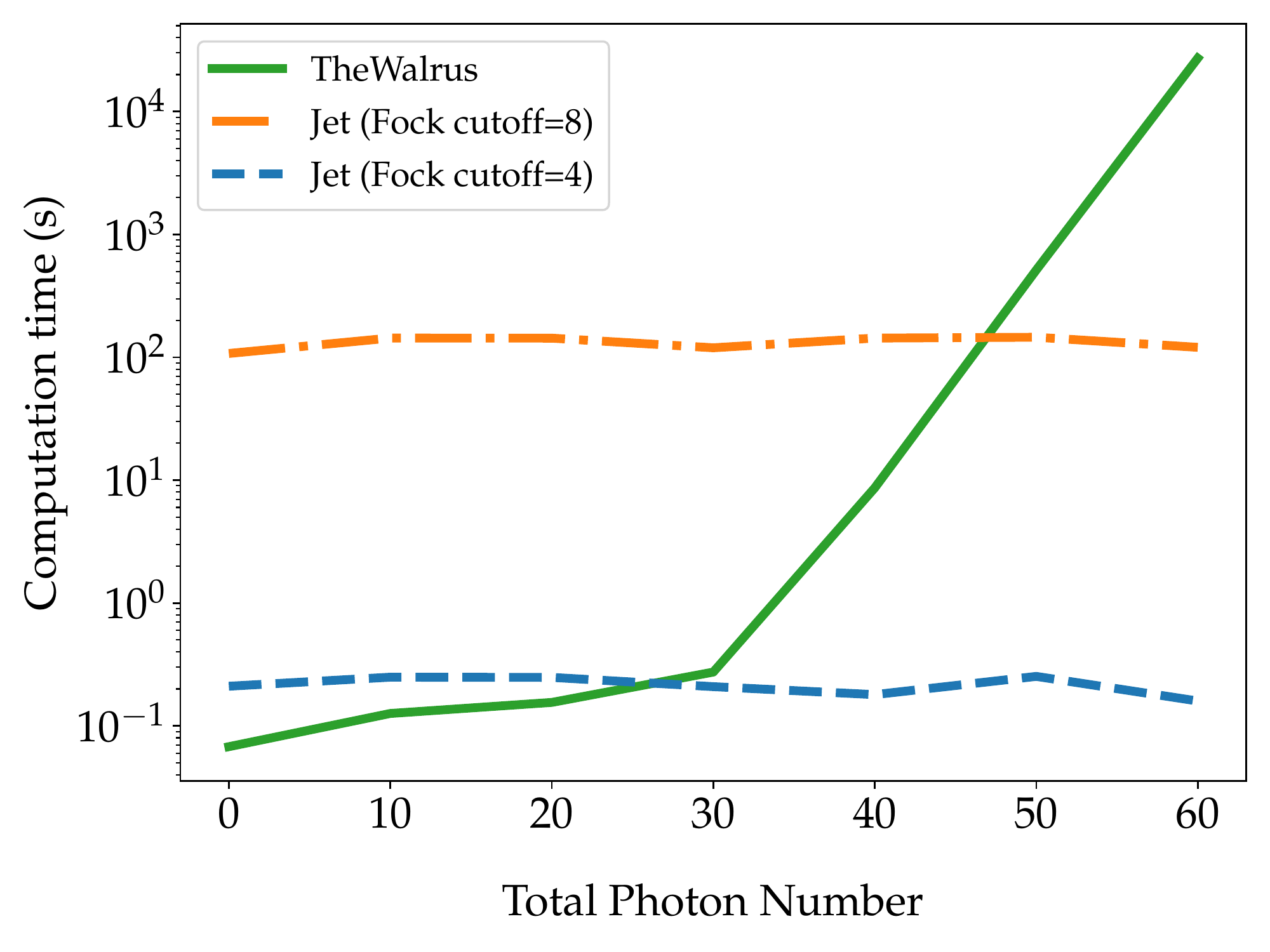}
    \caption{Jet and TheWalrus computation times for random amplitudes with different total photon numbers on a single node of the Niagara supercomputer. The amplitudes are generated by computing a sequence of random integers in the range [0,4) with the total sum of these integers representing the total photon number.~We compare times versus TheWalrus.~We compute results with Jet using single-precision complex numbers, whereas we use double-precision complex numbers in TheWalrus as this is natively supported. This would likely only slow down TheWalrus by a constant factor of 2, which would not explain the orders of magnitude difference in speed or scaling behaviour observed}
    \label{fig:gbs_runtimes}
\end{figure}

Finally, to date there are no known tensor-network simulators that can run GBS supremacy circuits.~We therefore benchmark against TheWalrus code, which computes amplitudes of GBS circuits by calculating a hafnian~\cite{gupt2019walrus}. Since the 3-dimensional random GBS circuits are computationally expensive, we benchmark on the 2-dimensional random GBS circuit GBS-88-m1 with a squeezing parameter $r=0.5$. For our tensor-network simulations we use a Fock basis cutoff of 4 and 8 for our tensors. We compute the same random amplitude using both TheWalrus and Jet with differing total photon number, $l$, and show the results in Fig.~\ref{fig:gbs_runtimes}.~The Fock states used for the amplitudes are generated by computing a sequence of random integers in the range $[0,4)$ with the total sum of these integers representing the total photon number, $l$.  

Figure~\ref{fig:gbs_runtimes} shows that TheWalrus computation time grows exponentially whereas Jet's computation time remains constant across total photon number at a fixed Fock cutoff. The constant nature of Jet's computation is due to the fact that at a fixed Fock cutoff, changing the total photon number only changes the values of some of the tensors in the network, but does not change the sizes of these tensors, or the contraction path. However, due to the Fock basis truncation, there is approximation error inherent in the Jet amplitude calculation. This approximation error becomes larger as the total photon number increases and can make Jet's computation differ from TheWalrus by orders of magnitude. Therefore it would be unfair to make concluding statements about the computation speed of TheWalrus versus Jet at fixed cutoff.  We leave a more thorough analysis of the approximation error and computation speed between Jet and TheWalrus to future work. For more details on the approximation error associated with tensor-network computation of optical quantum circuits, see Refs.~\cite{lubasch2018tensor,garcia2019simulating}.

\section{Conclusion}
\label{sec:conclusion}

The use of task-based parallelism for tensor-network simulation of quantum circuits provides multiple benefits which have not been utilized in previous simulators. Firstly, mapping the problem to a task-based framework allows for more parallelism without any extra work. Secondly, we can make use of shared work and potentially maximize it during the search for low-cost contraction orders and slices. Furthermore, this shared work can be pre-computed and stored on disk to be used for any slice or amplitude calculation for a given circuit.  Thirdly, a task-based method allows for better memory management as the tensors associated with tasks that no longer have any dependencies can be deleted on the fly, which in turn will allow for less slicing as the total memory needed is reduced. Lastly, task-based libraries allow for device agnostic computation, which will be a necessity as supercomputer nodes grow more heterogeneous. In this paper we also, for the first time, benchmark and compare Sycamore-53 supremacy circuits with GBS circuits.

For future work, there are many possible areas of further inquiry. As the
problems become more complex (e.g., sampling), the task graphs can become
dynamic and the shared work can grow dramatically. One could also extend our tasking to tensor networks with structure such as matrix product states (MPS), projected entangled pair states (PEPS), tree tensor networks and multi-scale entanglement renormalization ansatz tensor networks. We also did not thoroughly investigate methods for maximizing the shared work between multi-contractions through Eq. \ref{eq:task_based_amplitude_flops}. While we did use a basic greedy approach for maximizing the shared work between slices of the $m=10$ runs, this is very rudimentary and tends to give worse results as the tensor networks become larger. Maximizing shared work between multi-contraction is in general a very hard problem to develop heuristics for as it requires optimizing along several different areas over many different contraction trees. We also did not investigate using the SYCL portability layer Taskflow offers for other GPU vendors or more complicated CPU-GPU workflows that Taskflow supports.~Investigation of these additional directions is left for future work.~Lastly, it would be interesting to develop a framework for tensor-network simulations using distributed asynchronous tasking. Taskflow aims to support this in a future release and this may provide advantages over our current scheme which requires slicing to prevent memory overflow.
\\

\section{Acknowledgements}
The authors thank Johnnie Gray, Yohai Meiron, Fei Mao, Erik Spence and Ramses Van Zon for helpful discussions. The authors thank SOSCIP for their computational resources and financial support. We acknowledge the computational resources and support from SciNet. SciNet is funded by: the Canada Foundation for Innovation; the Government of Ontario; Ontario Research Fund - Research Excellence; and the University of Toronto. SOSCIP is funded by the Federal Economic Development Agency of Southern Ontario, IBM Canada Ltd. and Ontario academic member institutions.

\bibliographystyle{apsrev4-1}
\bibliography{jet}

\end{document}